\documentclass[a4paper,10pt,twoside]{cpc-hepnp}
\usepackage{upgreek,fancyhdr}
\usepackage{multicol}
\usepackage{graphicx}
\usepackage{booktabs}
\usepackage{amssymb,bm,mathrsfs,bbm,amscd}
\usepackage[tbtags]{amsmath}
\usepackage{lastpage}

\begin{document}

\fancyhead[c]{\small Chinese Physics C~~~Vol. xx, No. x (201x) xxxxxx}
\fancyfoot[C]{\small 010201-\thepage}

\footnotetext[0]{Received 31 June 2015}

\title{A Cockcroft-Walton PMT base with signal processing circuit\thanks{Supported by the West Light Foundation of The Chinese Academy of Sciences (29Y532050) and the Youth Innovation Promotion Association of the Chinese Academy of Sciences. }}

\author{%
      Yin Jun(Òü¿¡)$^{1,2;1)}$\email{yinjun@impcas.ac.cn}%
\quad Wang Yan-yu (ÍõÑåè¤)$^{1;2)}$\email{yanyu@impcas.ac.cn}%
\quad Zhang Ya-Peng(ÕÅÑÇÅô)$^{1)}$
\quad Ni Fa-fu(Äß·¢¸£)$^{1,2)}$ \\
\quad Zhang Peng-Ming (ÕÅÅôÃù)$^{1)}$
\quad Li Yao (ÀîÑþ)$^{1,2)}$
\quad Yuan Xiao-Hua (Ô¬Ïþ»ª)$^{1)}$
}
\maketitle

\address{%
$^1$Institute of Modern Physics, CAS, Lanzhou 730000, China\\
$^2$ University of Chinese Academy of Sciences, Beijing 100049, China\\
}

\begin{abstract}
Design a surface mount 14-PIN Cockcroft-Walton photomultiplier tube base for a muon detector, which provides both high voltage power supply and signal processing. The whole system, including the detector, adopts a +5V DC power input, and features as tiny size, low power-consumption and good portability, extremely well meeting the requirements of the power supply with a battery on a mobile workstation. Detailed descriptions and test results of a prototype are presented.
\end{abstract}



\footnotetext[0]{\hspace*{-3mm}\raisebox{0.3ex}{$\scriptstyle\copyright$}2013
Chinese Physical Society and the Institute of High Energy Physics
of the Chinese Academy of Sciences and the Institute
of Modern Physics of the Chinese Academy of Sciences and IOP Publishing Ltd}%


\begin{multicols}{2}

\section{Introduction}
Primary cosmic ray (mostly proton and $\alpha$ particles) from the universe hits earth continually. They strongly interact with nuclei of air gas at upper atmosphere, generating a particle shower at altitude of about 20 km. Secondary particles in the shower decay during the propagation and most of them end as long lived muons and neutrinos finally. On the ground, most of secondary charged cosmic-rays are muons. And average energy of the cosmic muon is about 4 GeV at the sea level that they can easily penetrate to underground with a depth of a few miles \cite{pdg}.

Typical energy loss of cosmic muon in the matter is about 2 MeV/cm. Mainly due to this effect, the cosmic ray contributes about 10\% of natural radiation dose. After a correlation between the cosmic ray flux and the cycles of biodiversity was observed \cite{rohde2005}, the cosmic ray is treated to be one of the driving forces of organic evolution on earth. The cosmic ray flux is also considered as one of possible reasons for the climate change, because the ionization from charged cosmic rays could enhance aerosol formation in the lower atmosphere \cite{Pedersen2012}. And the correlation between the cosmic ray flux with thickness of ozone \cite{Lu2009}, low cloud coverage \cite{Marsh2000} and temperature \cite{Dayananda2013} have been observed, but the final conclusion is still controversial.

Nevertheless, the cosmic muon exists everywhere at the surface of the earth, cosmic ray and their measurement widely connect to broad range of physics questions related to astrophysics, cosmology, particle physics, as well as organic evolution, which are quite inspirational for high school students especially. All these make the comic muon measurement to be an ideal science popularizing (SP) experiment.

In order to carry out the SP activities and also a future survey on the correlation between the cosmic muon flux and atmospheric parameters, e.g. temperature and pressure, we are planning to use the plastic scintillator and the photomultiplier tube (PMT) to build a portable and low-power consumption cosmic muon detecting device. Due to this reason, a low power consumption Cockcroft-Walton (CW) base for PMT with signal processing circuit was developed. The structure of this letter is organized as following: the first part is the introduction, the design of the CW base is described after it, then the performance tests of the CW base are presented and the last part is the conclusion of this letter.

\section{Circuit of Cockcroft-Walton base}

\subsection{CW circuit description}

CW circuit is an electric circuit that generates multiplying high DC voltages from a low AC or pulsing DC voltage input. This type of voltage multiplying circuit, consisting of a chain of diode rectifiers and capacitors, has no DC current flowing through the circuit, thus it has the advantage of dissipating much less power than a resistive voltage divider \cite{FEU85-3}. And it can provide a series of voltages with well-defined steps, which are well suitable to feed to a PMT¡¯s dynodes. Until now, CW base has been widely used as high voltage power supply for low power consumption device in high energy experiment \cite{INEEL}.

\begin{center}
\includegraphics[width=7cm]{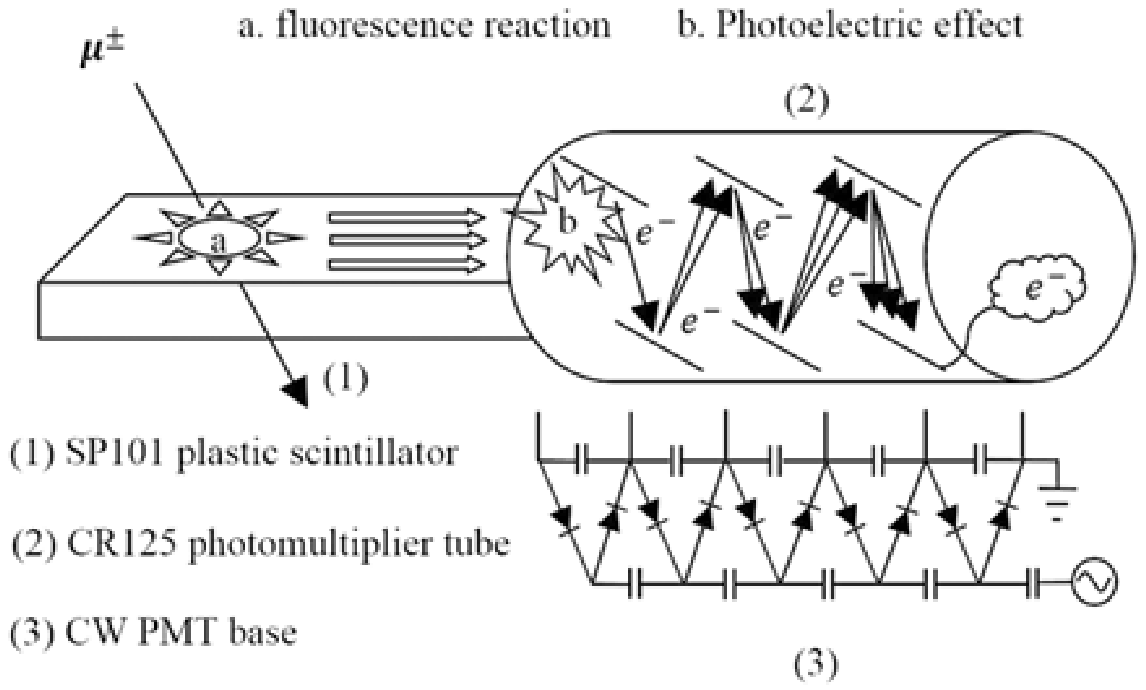}
\figcaption{\label{fig1}  Block diagram of the muon detector. }
\end{center}

We use a SP101 plastic scintillator as a probe to detect the cosmic muons, its scintillating light will be collected by a CR125 PMT from Hamamatsu \cite{CR125}. The PMT is connected to one end of the scintillator, other surfaces of the scintillator are wrapped by Tyvek paper and black tape as a light reflector and shading method. CR125 PMT is an 11 dynodes header-type PMT with effective area of 25mm$^2$, the maximum anode-to-Cathode voltage is 1500 V. The voltage distribution ratio from cathode (K), dynodes to anode (p) is 1:1:1:1:1:1:1:1:1:1:1:1 . A sketch of detection system including scintillator, PMT and CW base is shown in Fig. 1.

\begin{center}
\includegraphics[width=7cm]{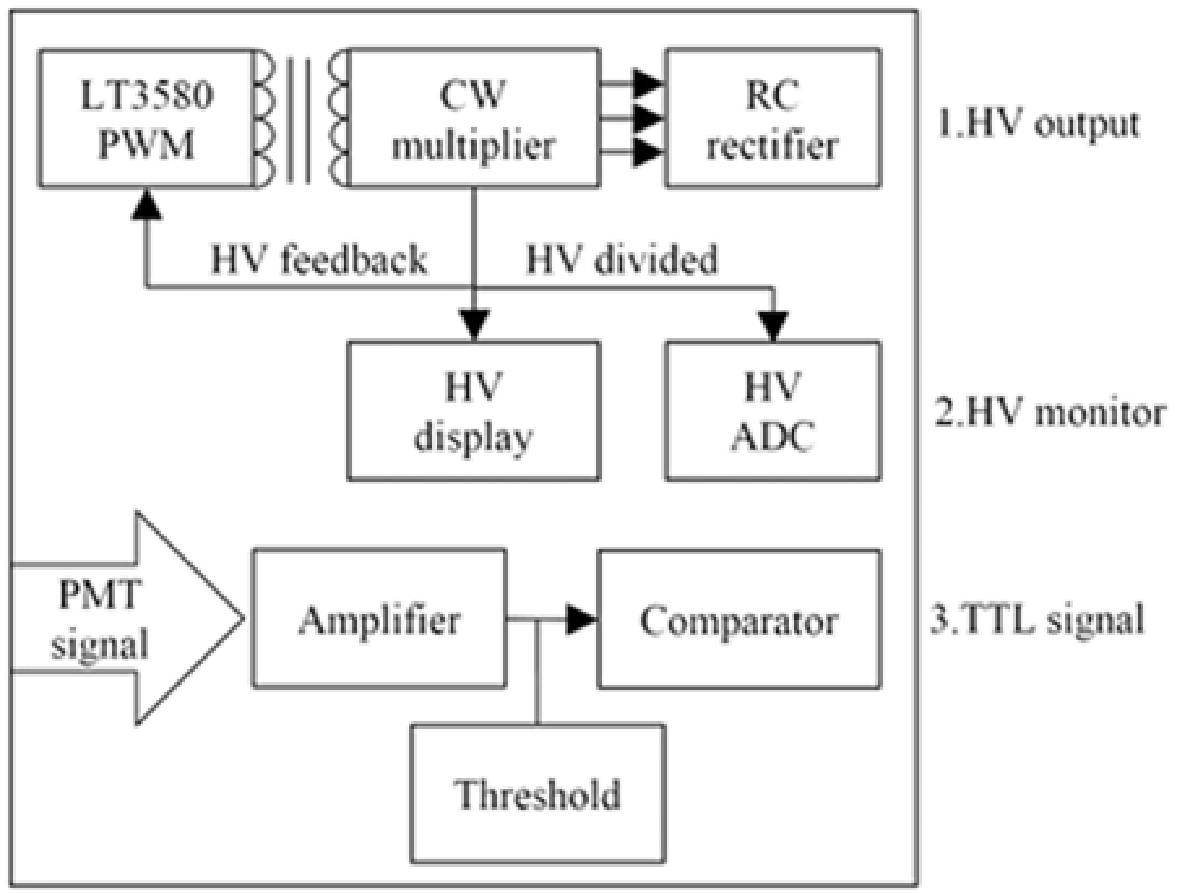}
\figcaption{\label{fig2}  Block diagram of the CW base. }
\end{center}

As shown in fig.2, on this base there are mainly three components as 1.high voltage power supply, 2.high voltage monitor and 3.tiny signal processing circuit. All these components are entirely contained on a printed circuit board, 3.3 cm in width and 9.3 cm in length. And the main parameters of the design are listed in Table 1.

\begin{center}
\tabcaption{ \label{tab1}  Main parameters of the CW base.}
\footnotesize
\begin{tabular*}{80mm}{l@{\extracolsep{\fill}}l}
\toprule
Photocathode voltage & -900 V$\sim$-1400 V $\pm$0.1\% \\
Long-time stability  & $\leq$0.15$\%$ \\
Maximum current & 15.8¦ÌA\\
Supply voltage   &  +5 V \\
Typical power consumption & 48mW(HV= -927V) \\
Maximum power consumption & 72 mW(HV= -1352V) \\
PMT signal pulse amplitude & -200mV$\sim$-500mV \\
PMT signal rising time & 5ns \\
PMT signal width &  30ns$\sim$50ns \\
TTL signal rising time & 10ns \\
TTL signal width & $\sim$1000ns \\
\bottomrule
\end{tabular*}
\end{center}

\subsection{High voltage power supply}

As a key component of the base (shown in fig.2), the high voltage power supply adopts a pulse width modulation (PWM) conversion chip LT3580 \cite{LT3580} to build its boost converter, and uses single-terminal fly-back to realize high voltage output. A 12-stages CW voltage multiplying rectifier supplies every dynode of PMT with divided voltage. The cut-off frequency of RC rectifier is 3.3Hz.

The LT3580 is a PWM DC/DC converter containing an internal 2A, 42V switch and can be configured as either a boost, SEPIC or inverting converter. It has an adjustable oscillator, set by a resistor from the RT pin to ground and uses a constant-frequency, current mode control scheme to provide excellent line and load regulation.

\begin{center}
\includegraphics[width=7cm]{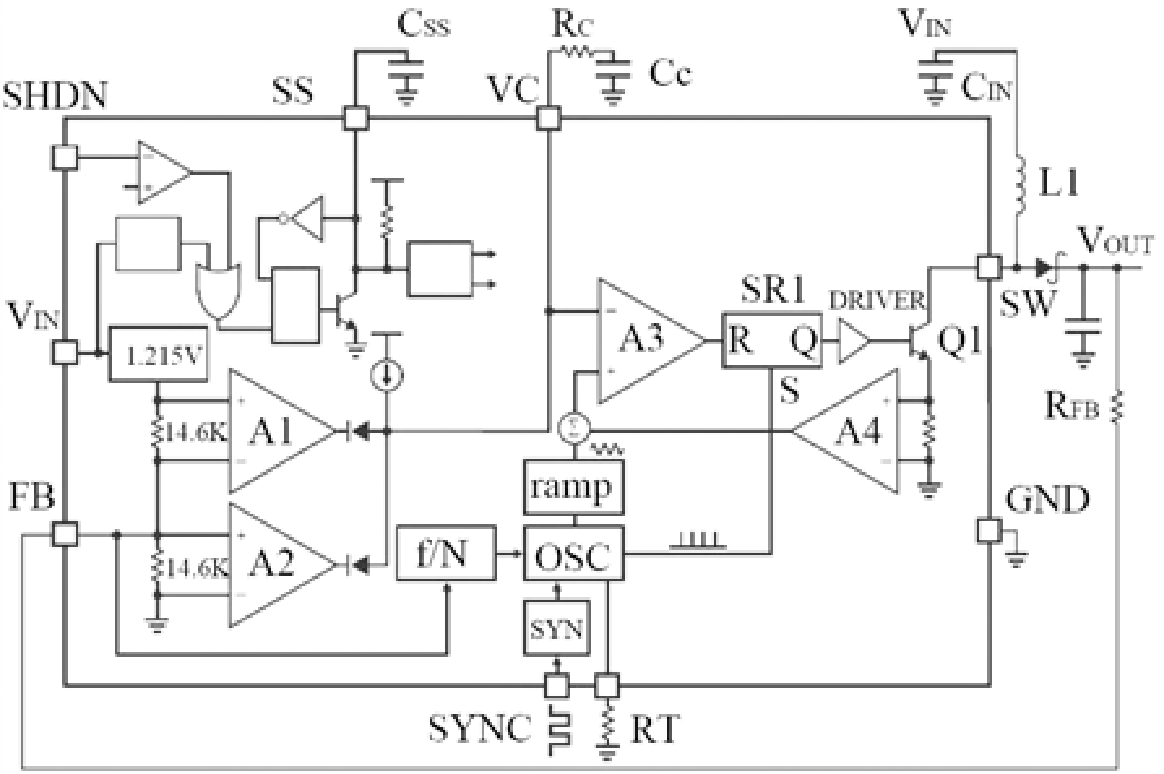}
\figcaption{\label{fig3}  Block diagram of the PMT base. }
\end{center}

As shown in fig.3, at the start of each oscillator cycle, the SR latch (SR1) is set, which turns on the power switch (Q1). The switch current flows through the internal current sense resistor generating a voltage proportional to the switch current. This voltage (amplified by A4) is added to a stabilizing ramp and the resulting sum is fed into the positive terminal of the PWM comparator A3. When this voltage exceeds the level at the negative input of A3, the SR latch is reset, turning off the power switch. The level at the negative input of A3 (VC pin) is set by the error amplifier A1 (or A2) and is simply an amplified version of the difference between the feedback voltage (FB pin) and the reference voltage (1.215V or 5mV depending on the configuration). In this manner, the error amplifier sets the correct peak current level to keep the output in regulation.

\begin{center}
\includegraphics[width=7cm]{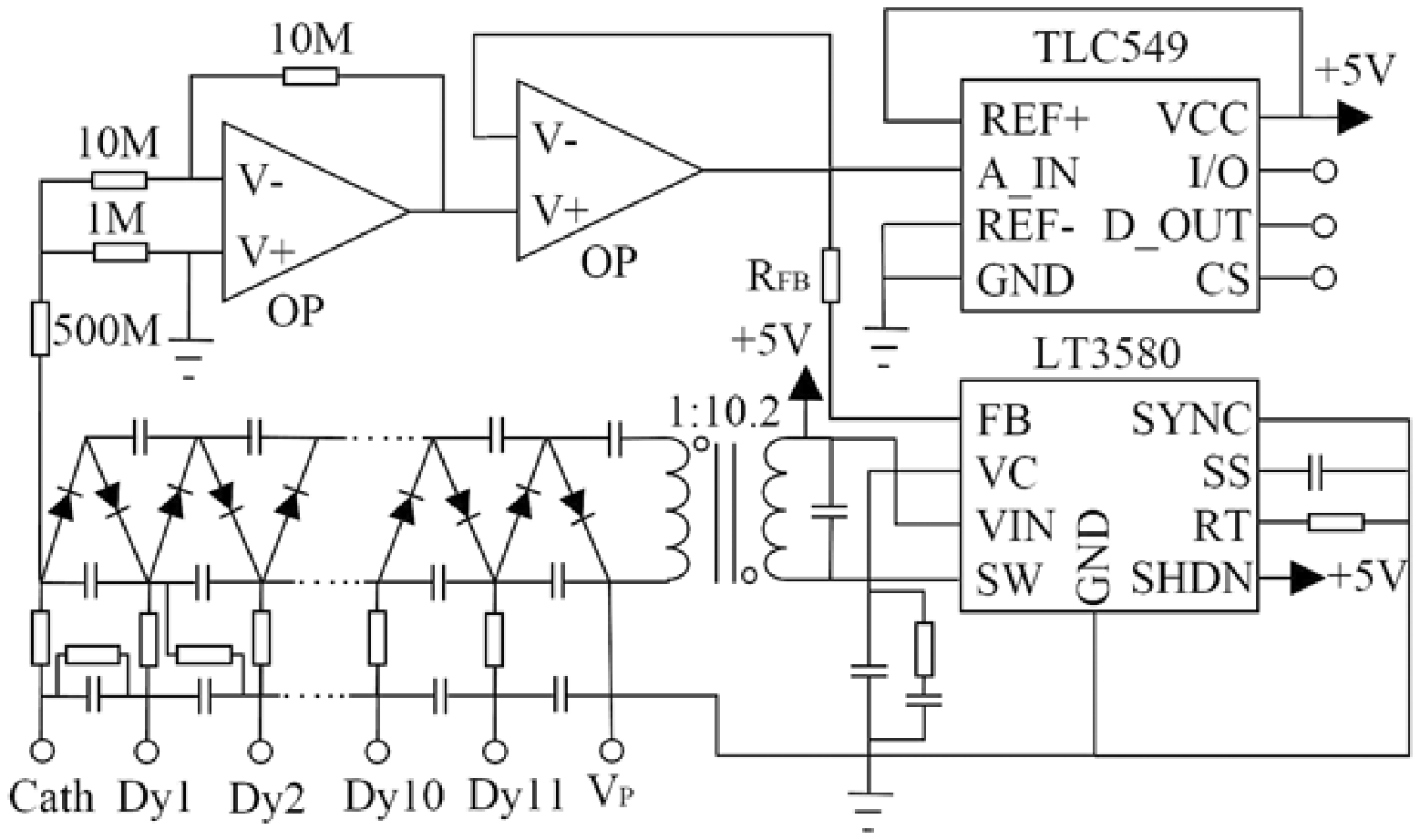}
\figcaption{\label{fig4} Block diagram of high voltage supply and monitor. }
\end{center}

As shown in fig.4, we adopt a few electronic components to configure LT3580 as a boost converter. A resistor connected from RT pin to GND is used to adjust the switching frequency at 500 KHz by equation (1). A series resistor-capacitor network with a capacitor in parallel is connected from the VC pin to GND to compensate the feedback loop. A tiny transformer (U2, TDK LDT565630T, turns ratio 1:10.2, sizes 5.8mm¡Á5.8mm¡Á3mm) is connected to the SW pin and then the pulsing DC output on switch are stepped up to about 100V to drive the CW multiplier. A capacitor in parallel with the transformer¡¯s primary windings will make the LC resonance frequency close to the switching frequency, then the transformer will get a higher voltage and the loss on switch can be reduced \cite{YaoFei}. The CW voltage multiplying rectifier has a recommended voltage divider ratio of 1:1:1:1:1:1:1:1:1:1:1:1 by the datasheet of CR125 PMT, and it will generates negative DC high voltages as $V_P$ pin is pulled down to GND.

In the left of CW, a circuit is placed to adjust the voltages of the electron focusing part of the PMT. This circuit provides filtering to the most sensitive part of the PMT for so called ¡®dark-count¡¯ pulse generation and transit time variations \cite{KM3NeT}. For there is a little current demand on the PMT($\leq$10uA), a first order low pass RC filter circuit is adopted to construct the dynode output filter and the cut-off frequency is 3.3Hz (R=1M$\Omega$£¬C=47nf) referring to equation(2). Since the feedback voltage (V$_{OUT}$ ) is positive, the FB pin shall be pulled down to 1.215V by adjusting the R$_{FB}$ resistor, then A2 becomes inactive and A1 performs the non-inverting amplification from FB to VC, as shown in fig.3. R$_{FB}$ is determined from equation (3).

\begin{equation}
\label{eq1}
SW_{frequency}(\text{MHz})=85387{\cdot}R_T^{-0.987} (K\Omega).
\end{equation}

\begin{equation}
\label{eq2}
f=\frac{1}{2{\pi}RC}.
\end{equation}

\begin{equation}
\label{eq3}
R_{FB}=\frac{V_{out}-1.215~V}{83.3 {\mu}A}.
\end{equation}

In addition, low ESR (equivalent series resistance) capacitors should be used in the CW voltage multiplying rectifier to minimize the output ripple voltages. Multilayer ceramic capacitors are an excellent choice, as they have an extremely low ESR and are available in very small packages. COG or NP0 dielectrics are preferred, as these materials have characteristics of temperature compensation and stable capacitance and dielectric loss. The capacitance changes $\pm$30ppm$\//$¡æ  at temperature varying from -55¡æ  to +125¡æ, less than ¡À0.3¦¤C with frequency. The capacitance drift or hysteresis is less than ¡À0.05\%. We choose NP0 type ceramic chip capacitors with 1206 case size and 10nf value as the multiplying and rectifying capacitors which could not only hold their capacitance at the desire output voltage, but also have low ESR with small output ripple. Assume that the output current is I (=10uA), the order of voltage multiplying is N (=12), the capacitance of capacitor is C (=10nf), and the switching frequency is f (=500 KHz), so a typical output voltage ripple is:
\begin{eqnarray}
  V_{rip} &=& \frac{(N+1){\times}N{\times}l}{4{\times}f{\times}c} ,\\
   &=&\frac{(12+1){\times}12{\times}10{\times}10^{-6}}{4{\times}500{\times}10^{3}{\times}10{\times}10^{-9}},  \\
   &=& 0.078 ~\text{V}.
\end{eqnarray}

\subsection{High voltage monitor}

The first part of high voltage monitor is a voltage divider composed of a 500 M$\Omega$ resistor in series with a 1 M$\Omega$ resistors. The voltage across the 1 M$\Omega$ resistor is a reduction factor of 1:501 of the actual voltage on cathode. To reduce circuit load, a follower circuit made by a voltage feedback type operational amplifier (OP), opa2350, is designed to receive the invert proportional voltage, as shown in fig.4, whose output voltage, provided as V$_{OUT}$ in formula (3), is on the one hand connected to FB pin through R$_{FB}$, and on the other hand is sent into an analog to digital conversion chip (ADC). The digital signals for voltage will be sent to a subsequent data acquisition card and shown in the upper interface through the network. Meanwhile, with the assistance of a digital voltmeter, the proportional voltage can be shown on the local as well.

\subsection{Signal process circuit}
The signal processing circuit receives weak signals from PMT, makes a ten-times amplification and compares the amplified signal with an adjustable threshold to gain the TTL output. The sampling resistor (R$_S$) acts on transforming a mass of electrons from the PMT output into a negative pulse with an amplitude of several millivolts and pulse width of tens of nanoseconds. A small value capacitor in parallel with the sampling resistor works on eliminating the pulse¡¯s tail oscillation to avoid generating fake signal pulses. A current feedback operarional amplifier (OPA2681) is adopted to construct a negative feedback amplifier with the amplification ratio of 1:10. Here is a power module(PUD0505-3K from DANUBE) accepting +5V input to provide ¡À5V power supply. Comparator contains a voltage comparison chip LT1711. The threshold is adjustable in -2.5V$\sim$0V. The TTL output is latched in order to accelerate signal turning speed, while also introducing hysteresis as shown in fig.5.
\begin{center}
\includegraphics[width=7cm]{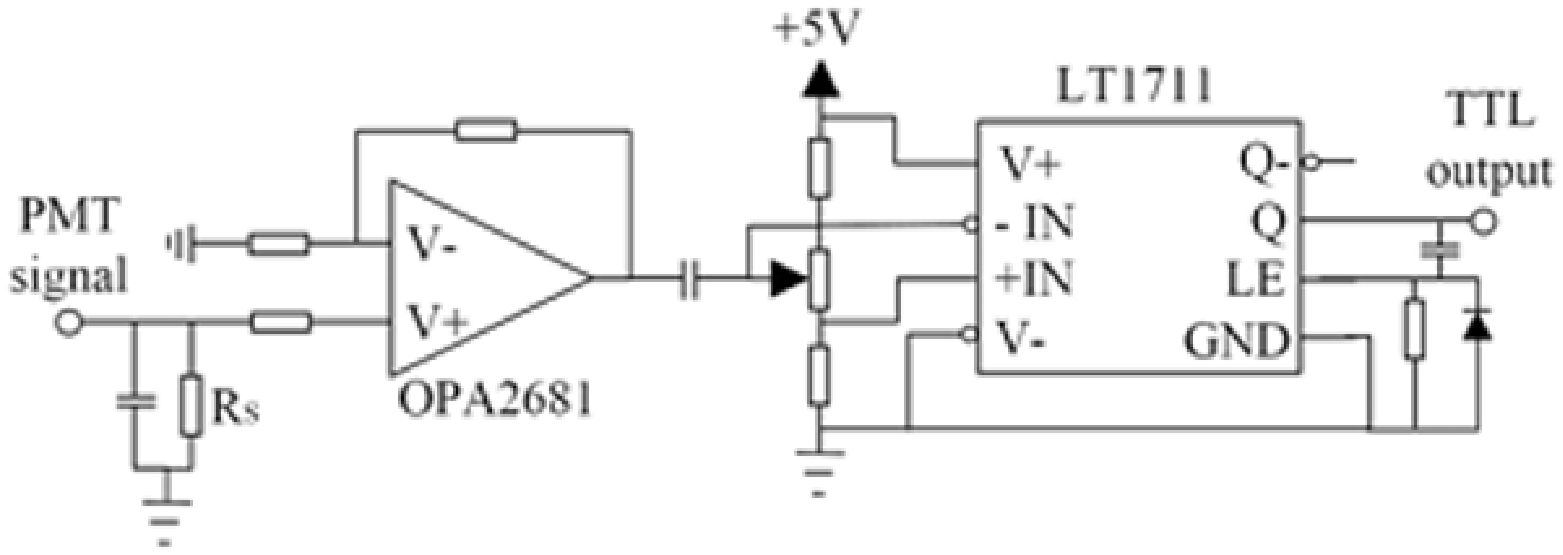}
\figcaption{\label{fig5} Block diagram of signal processing circuit. }
\end{center}

\section{Performance}

\subsection{Linearity and stability test}

We powered up the PMT with our CW base and measured the boost curves for dynode voltages. The measurement was carried out by using a 6-1$\//$2 digit precision multi-meter to measure and record one hundred data for each dynode voltage to calculate the average value and standard deviation. And the boost curves with a quite excellent linearity are shown in fig.6. The result shows that this high voltage power supply is able to provide PMT with up-to-standard working voltages. Meanwhile the result of a long-time stability test on cathode voltage was shown in fig.7, which indicates the cathode voltage has a very small drift ($\leq$0.15\%) during a long time.

\begin{center}
\includegraphics[width=7cm]{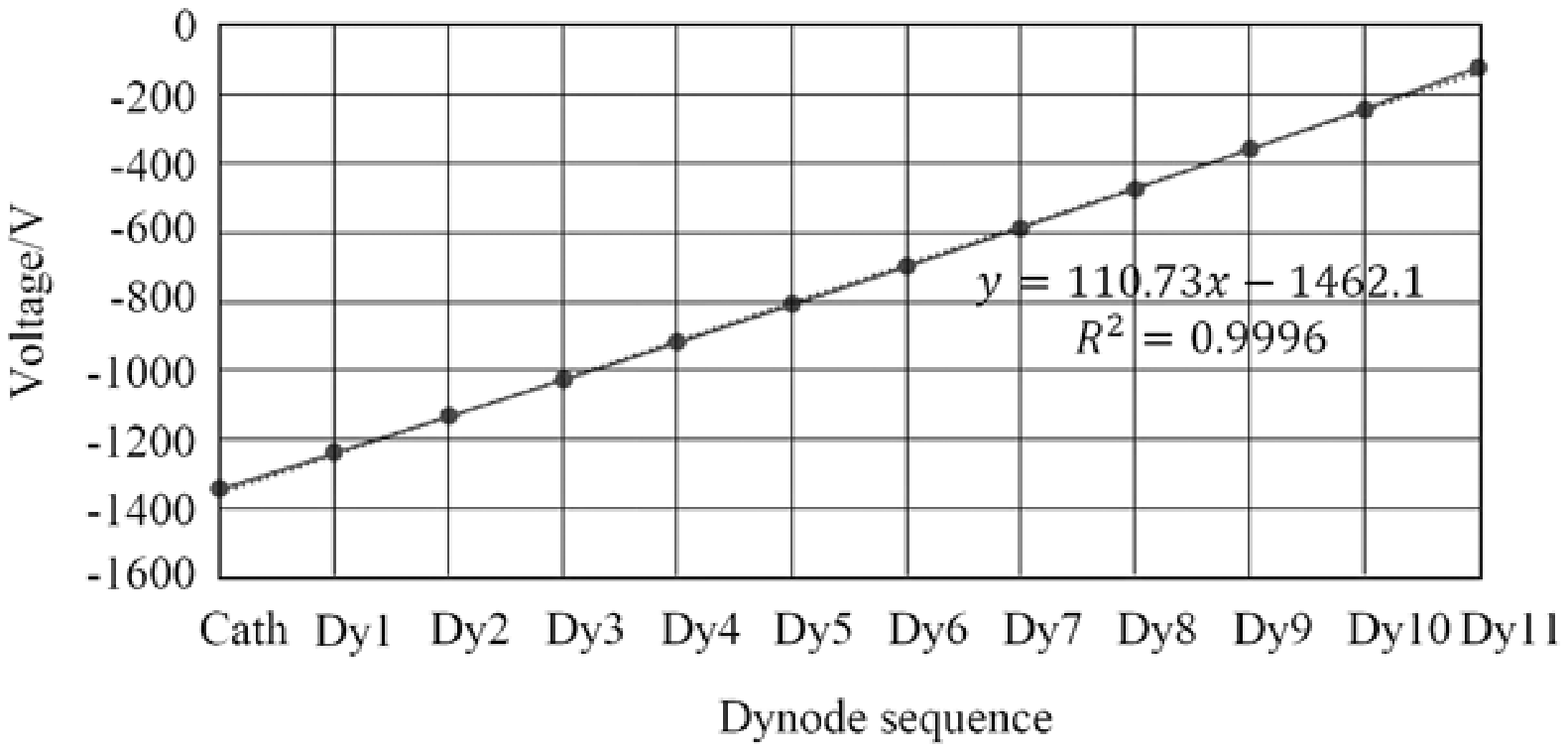}
\figcaption{\label{fig6} The boost curves of PMT base. }
\end{center}

\begin{center}
\includegraphics[width=7cm]{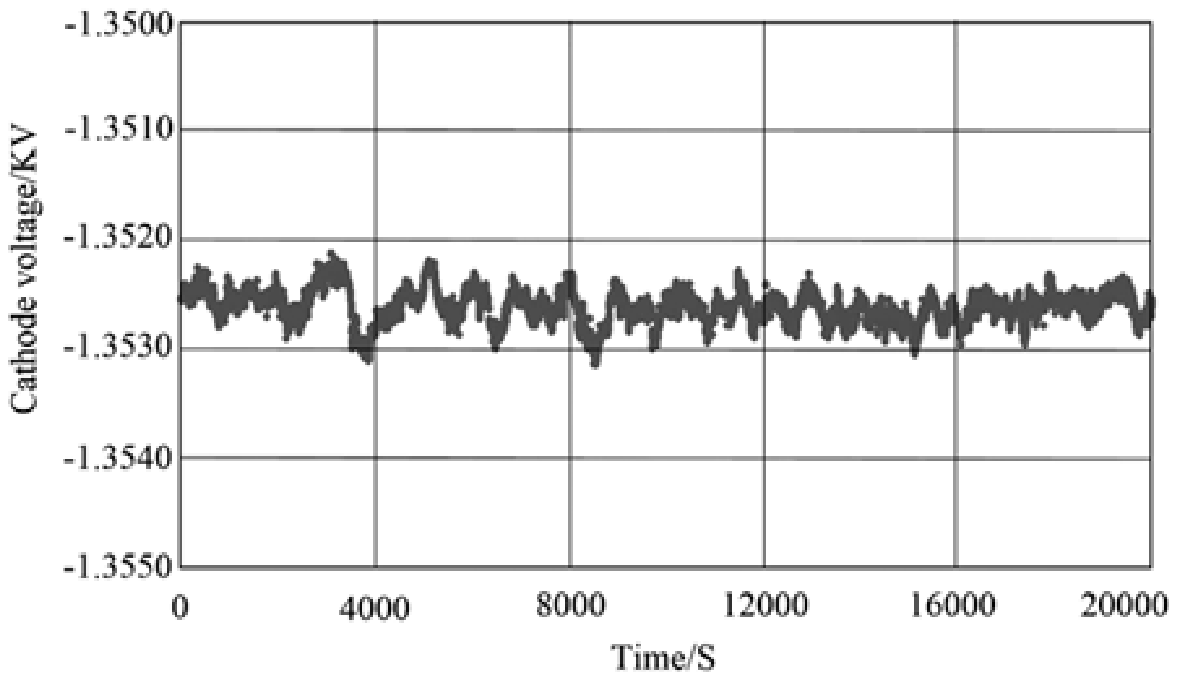}
\figcaption{\label{fig7} A long-time stability test on cathode voltage. }
\end{center}

\subsection{Plastic scintillator response and signal test}

We observed the PMT output electric signals on an oscilloscope after the PMTs were powered up. As shown in fig.8, the signals size -200mV$\sim$-500mV with 30ns$\sim$50ns (1ns=10$^{-9}$s) pulse width. The discriminator's latched output signal was shown in fig.9. The threshold value was set at around -150mV to remove the electronic noises.

\begin{center}
\includegraphics[width=7cm]{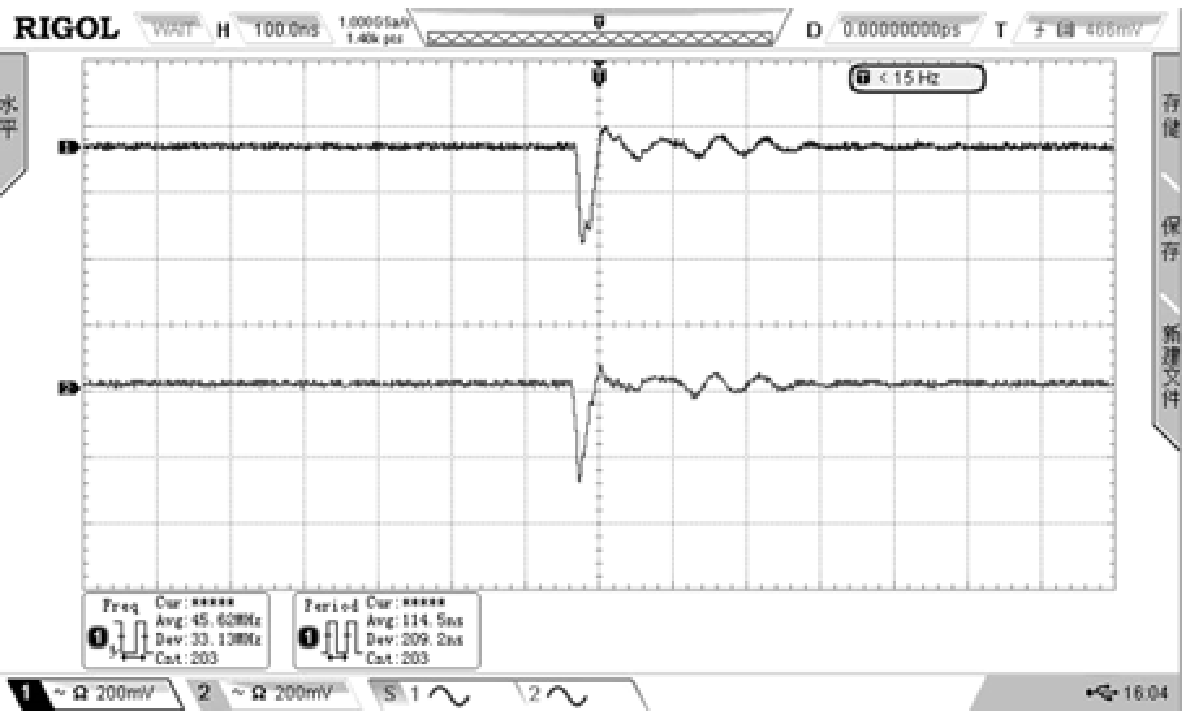}
\figcaption{\label{fig8} The plastic scintillator response. }
\end{center}

\begin{center}
\includegraphics[width=7cm]{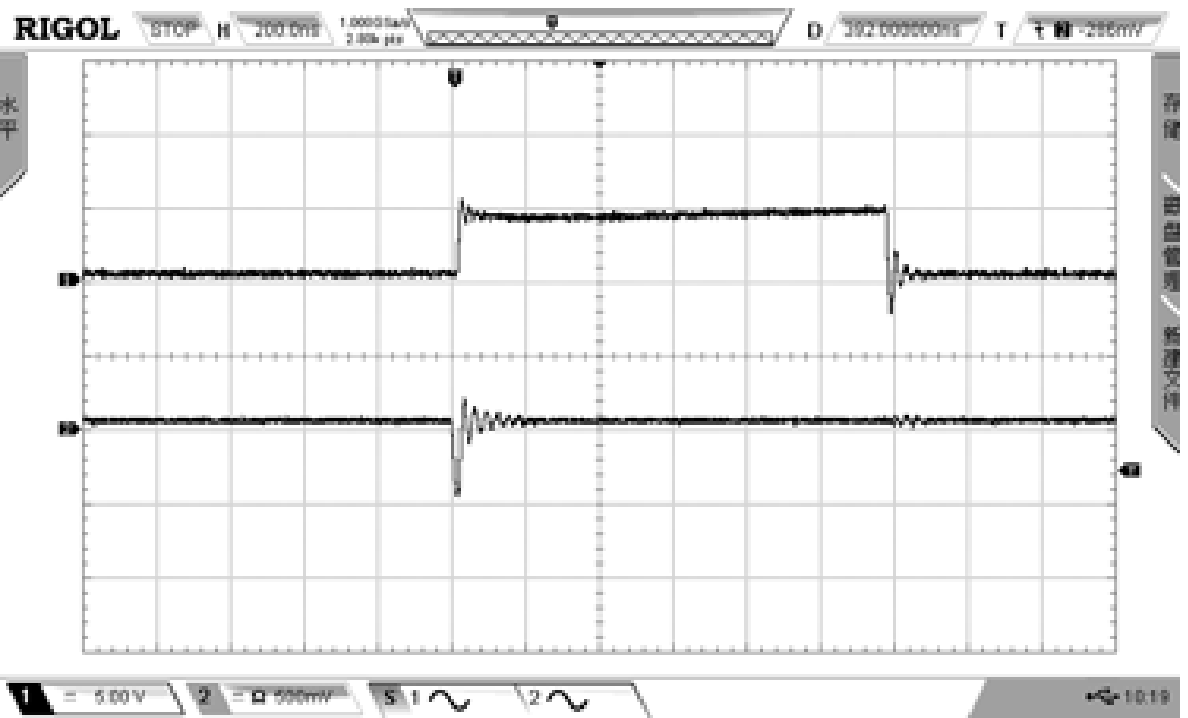}
\figcaption{\label{fig9} Discriminator latched output. }
\end{center}

\subsection{Implement in cosmic ray coincidence counting}

In order to gain a further inhibition of PMT noise, we received the coincidence of two signals from the plastic scintillators as a valid signal. The diagram of cosmic muon coincidence counting is shown in fig.10. With this platform we have obtained a preliminary measurement of the charged cosmic muon flux in more than two hours. The measurement consists of a real-time count and an accumulative count. As statistical results are shown in fig.11, the average flux was 24$\pm$2 per minute. Since our detectors area is about 0.01m$^2$, that is to say at every second in the vertical direction the number of charged cosmic ray through a 1m$^2$ area is about 40.
\begin{center}
\includegraphics[width=7cm]{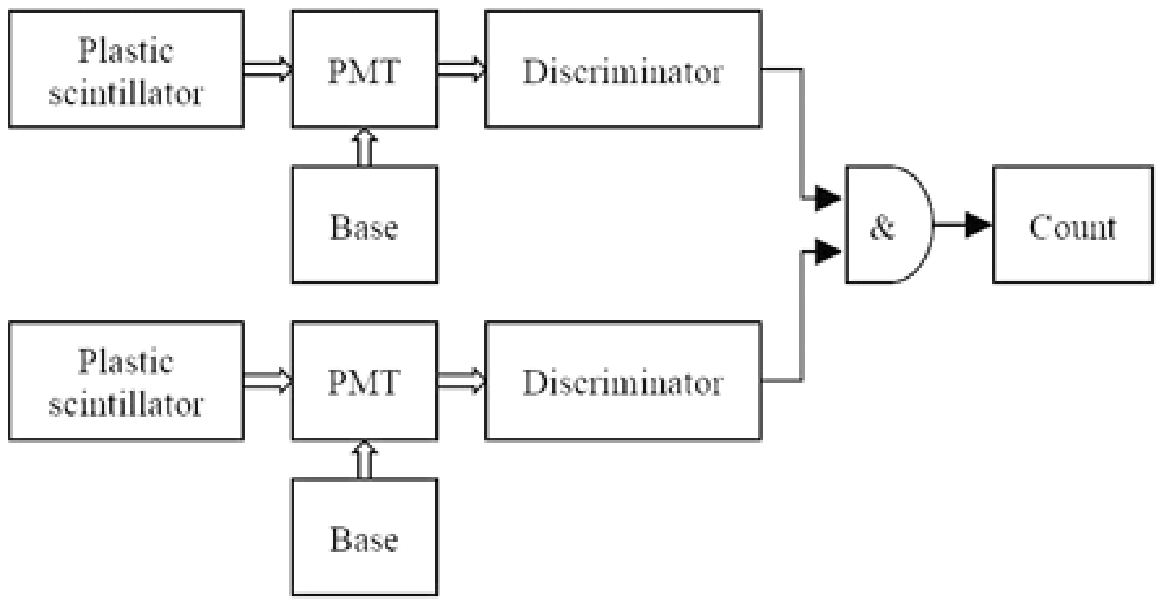}
\figcaption{\label{fig10} Diagram of cosmic muon coincidence counting. }
\end{center}
\begin{center}
\includegraphics[width=7cm]{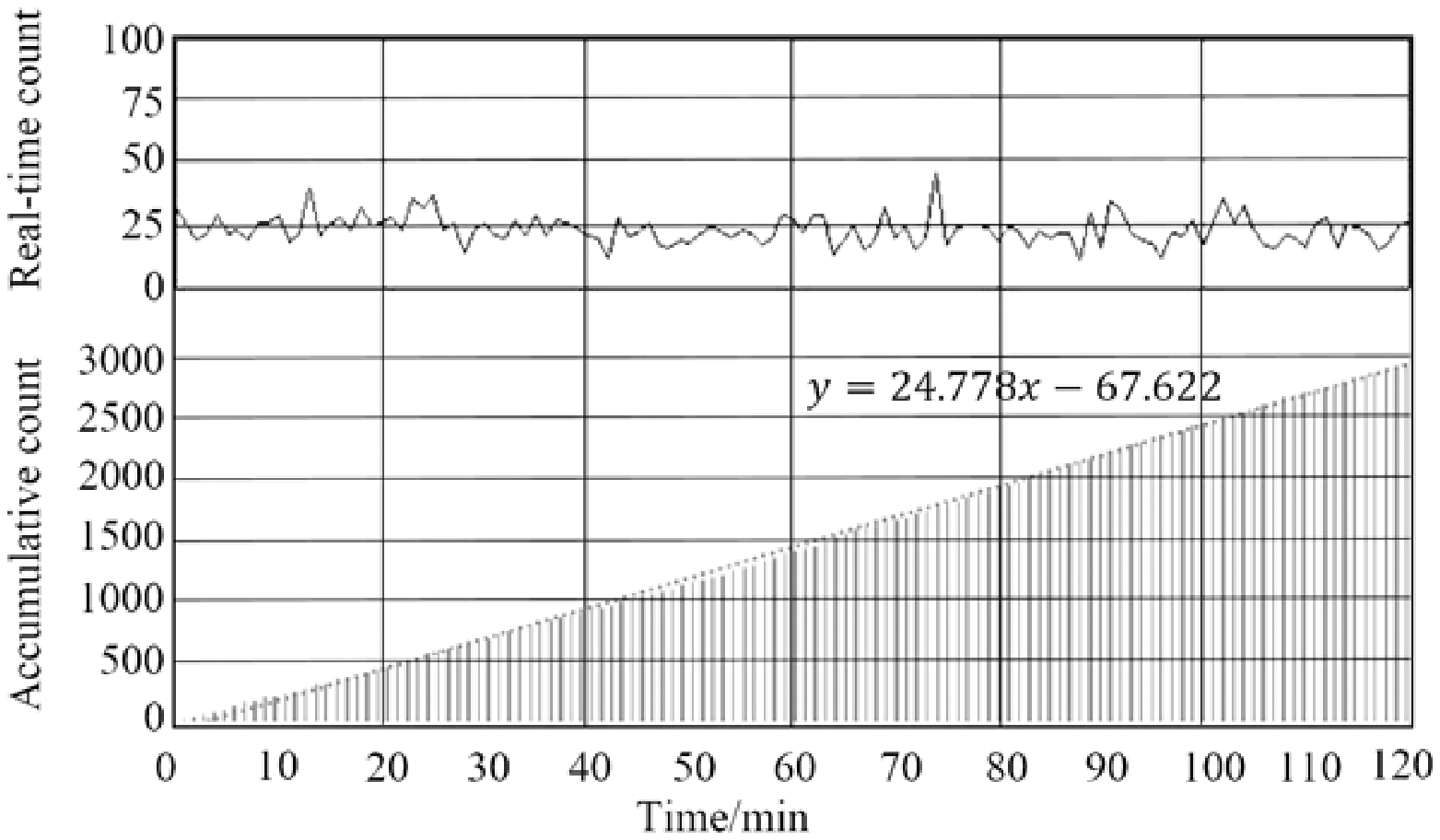}
\figcaption{\label{fig11} Result of cosmic muon coincidence counting. }
\end{center}

\section{Conclusion}
This CW base for PMT adopts a +5V battery input to generate a -900V$\sim$-1400V adjustable linear output and its long-time voltage stability is $\leq$0.15$\%$. The entire module features as tiny size, good portability, low power consumption and is able to provide effective measurement signals. It extremely well meets the requirements of the power supply for PMTs to carry out the SP activities and a future survey on the correlation between the cosmic muon flux and atmospheric parameters on the mobile workstation.

\end{multicols}

\vspace{-1mm}
\centerline{\rule{80mm}{0.1pt}}
\vspace{2mm}

\begin{multicols}{2}

\end{multicols}

\clearpage

\end{document}